\documentclass[10pt, a4paper, UKenglish]{article}
\usepackage{siunitx}
\usepackage{graphicx}
\def\Title#1{\begin{center} {\Large #1 } \end{center}}
\def\Author#1{\begin{center}{ \sc #1} \end{center}}
\def\Address#1{\begin{center}{ \it #1} \end{center}}

\newcommand\pubblock{\rightline{\begin{tabular}{l} Proceedings of the CTD/WIT 2019\\ \pubnumber\\
         \pubdate  \end{tabular}}}

\newenvironment{Abstract}{\begin{quotation} \begin{center} 
             \large ABSTRACT \end{center}\bigskip 
      \begin{center}\begin{large}}{\end{large}\end{center} \end{quotation}}

\newenvironment{Presented}{\begin{quotation} \begin{center} 
             PRESENTED AT\end{center}\bigskip 
      \begin{center}\begin{large}}{\end{large}\end{center} \end{quotation}}

\def\beq{\begin{equation}}
\def\eeq#1{\label{#1}\end{equation}}
\def\eeqn{\end{equation}}

\def\beqa{\begin{eqnarray}}
\def\eeqa#1{\label{#1}\end{eqnarray}}
\def\eeqan{\end{eqnarray}}

\def\Dslash{\not{\hbox{\kern-4pt $D$}}}
\def\dslash{\not{\hbox{\kern-2pt $\del$}}}

\def\Pl{{\mbox{\scriptsize Pl}}}

\def\msb{{\bar{\ssstyle M \kern -1pt S}}}

\usepackage{color}
\usepackage{lineno}
\usepackage{subfig}
\usepackage{hyperref}
\usepackage{xspace}
\usepackage{multirow}
\usepackage{heppennames2}
\usepackage{CLICdp_definitions}

\DeclareSIUnit\mrad{\milli\radian}
\DeclareSIUnit\rad{\radian}

\newcommand\pubnumber{CLICdp-Conf-2019-005}

\newcommand\pubdate{\today}

\def\affiliation{
On behalf of the CLICdp collaboration, \\
Experimental Physics Department at CERN \\
Geneva, Switzerland}

\newcommand{\conference}{Connecting the Dots and Workshop on Intelligent Trackers (CTD/WIT 2019)\\
Instituto de F\'isica Corpuscular (IFIC), Valencia, Spain\\ 
April 2-5, 2019}

\usepackage{fancyhdr}
\definecolor{mygrey}{RGB}{105,105,105}
\begin{document}

% uncomment the following line for adding line numbers
% \linenumbers

% large size for the first page
\large
\begin{titlepage}
\pubblock

%% Title 
\vfill
\Title{New developments in conformal tracking for the CLIC detector}
\vfill

%  if you need to add the support use this, fill the \support definition above. 
%  \Author{FIRSTNAME LASTNAME \support}
\Author{Erica Brondolin}
\Address{\affiliation}
\vfill

\begin{Abstract}
Conformal tracking is an innovative track finding strategy adopted for the detector at
the Compact Linear Collider (CLIC), a proposed future electron--positron collider.
It features a pattern recognition in a conformal-mapped plane using the cellular automaton
algorithm to reconstruct the trajectory of charged particles in a magnetic field.
The efficiency and robustness of the algorithm are validated using full-simulation studies
in the challenging beam-induced background conditions expected for the 3 TeV stage of the CLIC collider.
The tracking performance requirements, set by the ambitious CLIC physics programme, have been shown to be met.
Moreover, thanks to its flexibility and geometry-agnostic nature, 
this algorithm was also shown to be easily adaptable to different detector designs and beam conditions.
\end{Abstract}

\vfill

% DO NOT CHANGE!!!
\begin{Presented}
\conference
\end{Presented}
\vfill
\end{titlepage}
\def\thefootnote{\fnsymbol{footnote}}
\setcounter{footnote}{0}
%

% normal size for the rest
\normalsize 

%% Your paper should be entered below. 

\section{Introduction}
\label{sec:introduction}

The next generation of tracking detectors foreseen for future electron--positron colliders, 
such as the Compact Linear Collider (CLIC),
will feature ultra-low-mass systems and provide extremely high position accuracy
to fulfil the requirements imposed by the physics programme
and the challenging beam-induced background conditions.
In order to fully exploit these features in track reconstruction, a new track finding technique 
was developed by the CLICdp collaboration that merges the two concepts of conformal mapping and cellular automaton.
We refer to this technique as \emph{conformal tracking}. 
Thanks to its flexibility and robustness, the new algorithm can also be
adapted easily to different detector designs and geometries.

The goal of this paper is to describe the conformal tracking algorithm, its implementation as well 
as its performance in the track reconstruction used in the CLIC detector (CLICdet).
Section~\ref{sec:clic} introduces the future CLIC collider with
particular emphasis on the physics requirements and the beam-induced background conditions including
the resulting challenges in terms of track performance.
It concludes with a brief description of the tracking system for CLICdet.
In Section~\ref{sec:simulation} the event simulation and software framework are presented.
Section~\ref{sec:track_reco} presents the track reconstruction process at CLIC.
The tracking performance in terms of efficiency, fake rate, track resolution and CPU time
is shown in Section~\ref{sec:performance}, while conclusions are presented in Section~\ref{sec:summary}.

\section{Tracking challenges at CLIC}
\label{sec:clic}

CLIC is a proposed \epem collider operating in three stages at 
centre-of-mass energies of \SI{380}{\GeV}, \SI{1.5}{\TeV} and \SI{3}{\TeV},
each stage lasting $7-8$ years~\cite{cdrvol2, cdrupdate, clic_summaryReport}. 
CLIC's main goals are to measure the properties of the top quark and the Higgs boson with high precision
and to search for physics beyond the Standard Model.

In order to reach its design luminosity of $1.5-6\times 10^{34}$cm$^{-2}$s$^{-1}$, 
CLIC will operate with very small bunch sizes 
%(less than  \SI{150}{\nano\meter} in $x$ and 
%\SI{3}{\nano\meter} in $y$ and less than \SI{100}{\micron} along the beam)
~\cite{cdrvol2, cdrupdate, clic_DT}. 
CLIC's bunch sizes lead to strong beamstrahlung radiation from the electron
and positron bunches in the field of the opposite beam. The interactions involving
%The creation of the beamstrahlung photons
%reduces the available centre-of-mass energy of the \epem collisions and interactions involving 
beamstrahlung photons result in two main types of background, incoherent \epem pairs and \gghadron{} events.
Both backgrounds are expected to affect the tracking performance in terms of detector occupancy.
The incoherent \epem pairs are mostly emitted in the forward direction
and constrain the inner detector radius to $\geq\SI{31}{\mm}$ in the central region of the detector.
The \gghadron{} events extend to larger angles with respect to the beam line, thus impacting the physics measurements. 
%For this reason, only this background is included 
%at the time of this paper as first step to study the track reconstruction robustness.
At the \SI{3}{\TeV} stage, each bunch train contains 312 bunches, separated by \SI{0.5}{\ns}. The
bunch trains are separated by \SI{20}{\ms}.
%This beam structure allows for power pulsing of the CLICdet and a trigger-less readout.

An accurate vertex reconstruction is needed to obtain an accurate and efficient identification 
of secondary vertices from heavy-quark flavour hadrons and tau-leptons.
This requirement translates to a transverse impact-parameter resolution 
at the level of $\sigma_{d_0}^2 = (\SI{5}{\um})^2 + (\SI{15}{\um\GeV})^2/(p^2\sin^{3}\theta)$
and thus to a single point resolution of $3~\micron$ for the vertex detector, 
the innermost detector of CLICdet~\cite{clic_DT, CLICdet}.
The vertex detector is made of $25\times25~\si{\micron}^2$ pixels which 
are arranged in three cylindrical double-layers in the central region of CLICdet and
in double-layer petals in a spiral arrangement in the forward direction
allowing for efficient air cooling of the entire vertex detector.

In terms of track reconstruction, the scientific goals of CLIC place also a demanding requirement
on the transverse momentum resolution for high-momentum tracks in the central detector region, 
which must be at the level of $\sigma_{\pT}/\pT^2 \leq \SI{2e-5}{\per\GeV}$.
Optimization studies show that a tracker with a radius of \SI{1.5}{\meter}
immersed in a magnetic field of 4~T with a single point resolution 
smaller than $7~\micron$ will achieve the required resolution~\cite{clic_DT}.
Therefore, a tracking detector made of silicon elongated pixel layers is placed 
around the vertex detector. 

\begin{figure}[!htb]
  \centering
  \subfloat[]{\includegraphics[width=.55\textwidth]{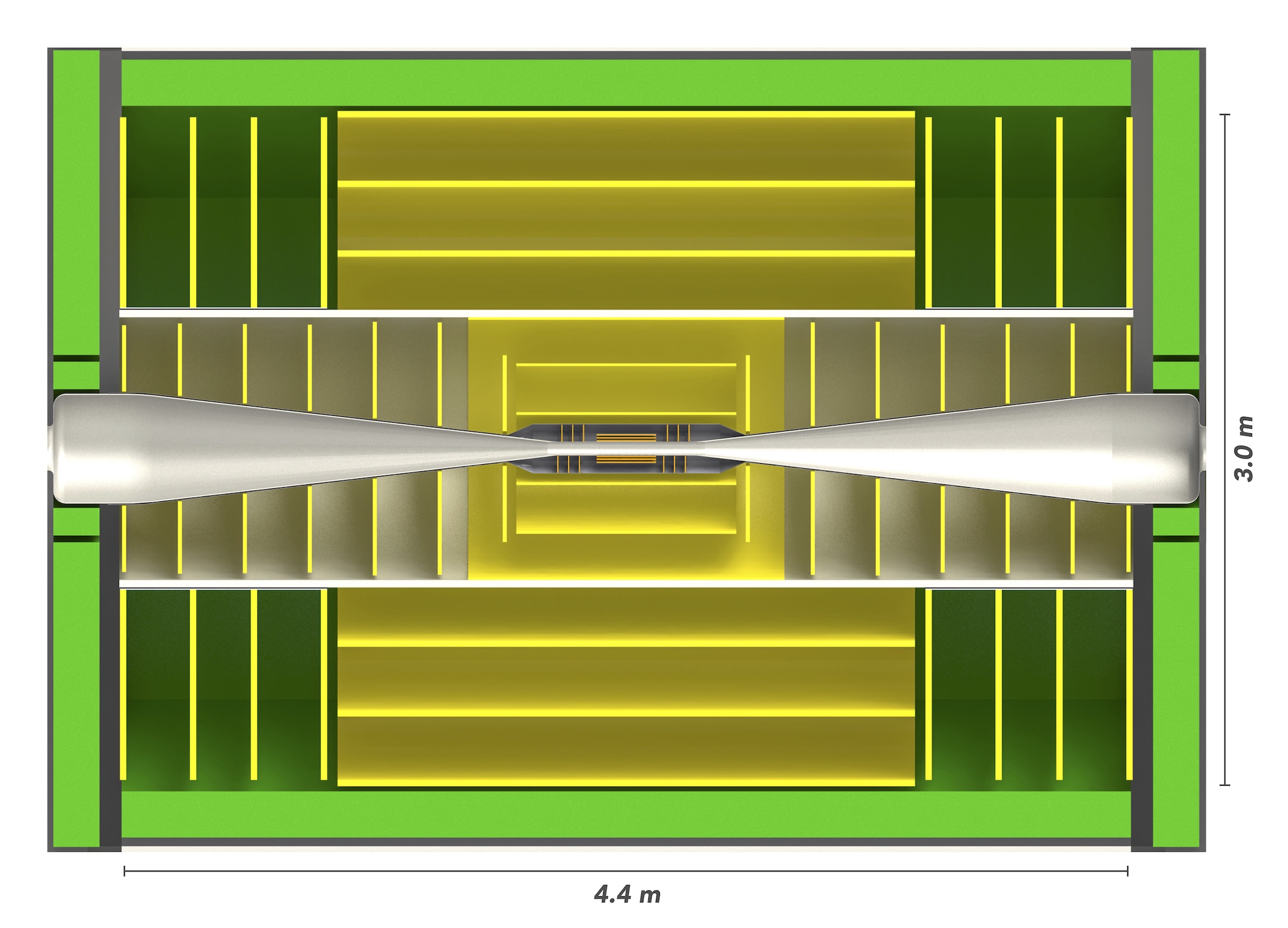}}
  \subfloat[]{\includegraphics[width=.45\textwidth]{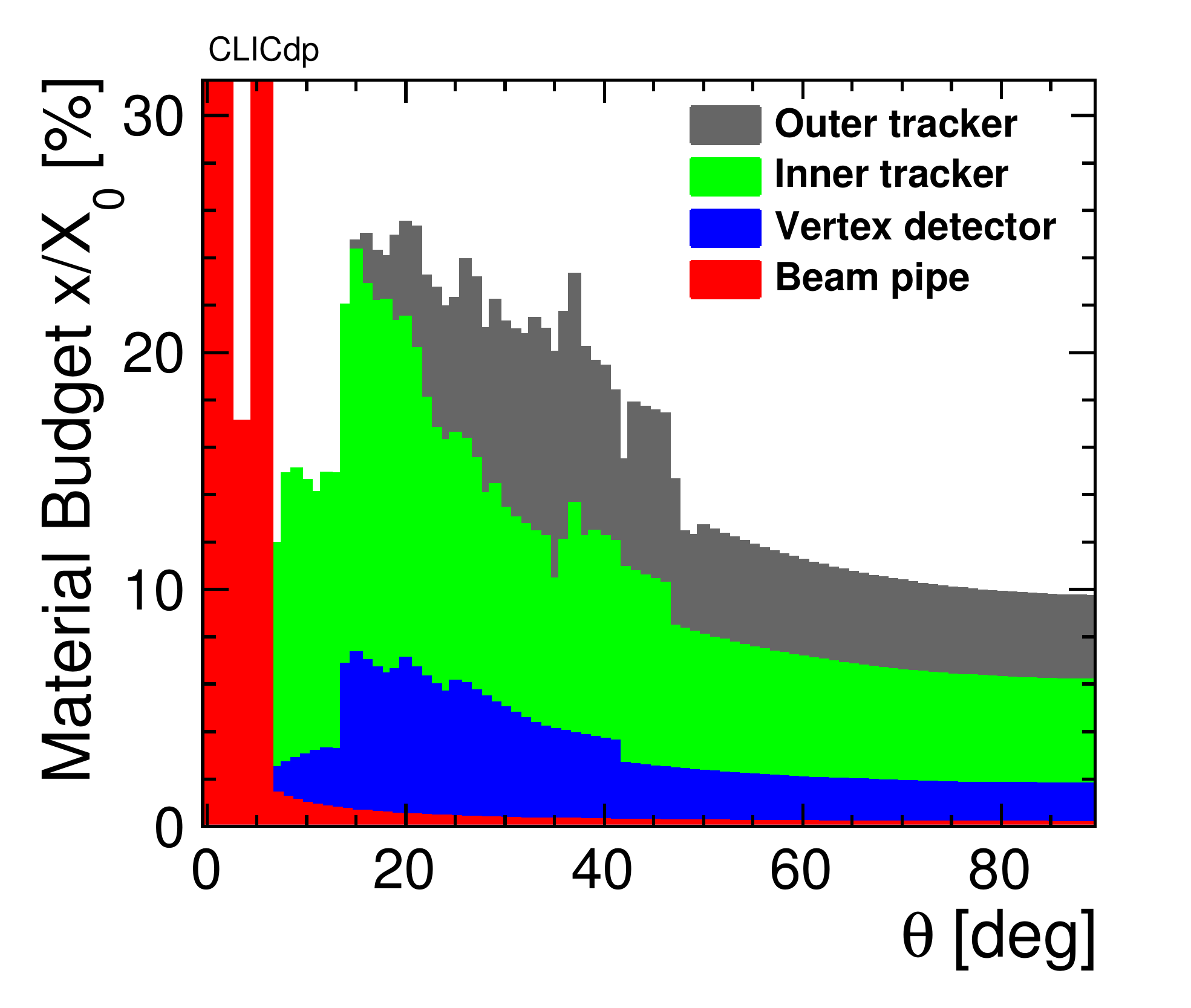}}
  \caption{(a) Tracking system layout in CLICdet~\cite{clic_DT}. The vertex-detector layers are shown in orange, while 
  the main tracker layers are depicted in yellow. The tracker is surrounded by the electromagnetic calorimeter, depicted in bright green.
  (b) The total thickness $x$ of the tracker material
  as a function of the polar angle and averaged over azimuthal angles, expressed in units of radiation length
  $X_0$~\cite{CLICperf}.}
  \label{fig:CLICdet}
\end{figure}

To fulfil the stringent requirements from the physics programme, 
the material budget of the CLICdet tracker must be limited to $0.2\%$ radiation length ($X_0$) per single layer
in the vertex detector and $1-2\% X_0$ per layer in the tracker. 
This ultra-light tracker with an extremely precise %hit-timing and 
hit-position resolution
is placed in the innermost part of CLICdet as shown in Figure~\ref{fig:CLICdet}(a).
Its total amount of material expressed in term of radiation length $X_0$ is shown in~\ref{fig:CLICdet}(b)
and corresponds to about $4\%$ nuclear interaction lengths in the central detector region.

\section{Event simulation and reconstruction}
\label{sec:simulation}

The performance of the track reconstruction algorithm used in CLICdet
is estimated with full simulation studies. The full tracker geometry 
including the support material, cables and cooling
is described with the DD4hep software~\cite{frank15:ddg4} 
and simulated in \geant~\cite{Agostinelli2003}
and includes a homogeneous solenoid field of \SI{4}{T}.

To assess the tracking performance, two types of events are simulated and reconstructed.
The first sample contains single muons and possible 
secondary particles produced by interaction with material.
The second sample represents a more complex scenario where a jet topology event,
an \epem$\rightarrow$ \ttbar event at \SI{3}{\TeV} centre-of-mass energy,
is generated with \whizard\cite{Kilian:2007gr} and the \gghadron background 
expected at the \SI{3}{\TeV} CLIC energy stage is superimposed~\cite{Schade:1443537}.
The \gghadron background overlaid corresponds to a total of 30 bunch crossings (BX),
10 bunch crossings before and 20 bunch crossings after the physics event.
The \ttbar event at \SI{3}{\TeV} produces on average about $90$ tracks,
which increases to about $550$ tracks when the \gghadron  background is overlaid.
%A hit map of a single \SI{3}{\TeV} centre-of-mass energy \ttbar event simulated in CLICdet 
%is shown in Figure~\ref{fig:hitsDisplay}.

The CLIC track reconstruction software is implemented in the linear collider Marlin framework~\cite{MarlinLCCD}
and interfaced with the geometry using DD4hep~\cite{sailer17:ddrec}. 
Large simulation and reconstruction samples were produced with the iLCDirac grid production system~\cite{ilcdirac13}.

%\begin{figure}[!htb]
%  \centering
%  \subfloat[]{\includegraphics[width=0.45\linewidth]{figures/detectorAndHits_xy}}
%  \subfloat[]{\includegraphics[width=0.55\linewidth]{figures/detectorAndHits_yz}}
%  \caption{Reconstructed hits produced in the CLICdet tracking system by a \SI{3}{\TeV} 
%  centre-of-mass energy \ttbar event simulated with 30 BX overlaid
%  shown in the $x-y$ (a) as well as in $y-z$ (b) plane.}
%  \label{fig:hitsDisplay}
%\end{figure}

\section{Track reconstruction}
\label{sec:track_reco}

The main goal of the track reconstruction process is to estimate the momentum and position parameters of
charged particles crossing the detector in a magnetic field.
The first step of the track reconstruction is to create \emph{reconstructed} hits. 
In CLICdet, the local position of the \emph{simulated} hits from charged particles crossing the sensitive layers of the detector 
are smeared with a Gaussian distribution with a width equal to the single point resolution of the subdetector.
This procedure is done for all simulated hits.
The reconstructed hits are used as input for the subsequent steps of the track reconstruction.

The CLIC track reconstruction can be divided in three main blocks:
the track finding with the conformal tracking algorithm, the track fitting using a Kalman filter and smoother,
and the track selection. The final output of the track reconstruction
are tracks which include not only the hits they are made of, but also the information of the estimated track parameters.

\begin{figure}[bt]
  \centering
  \subfloat[]{\includegraphics[width=0.48\linewidth]{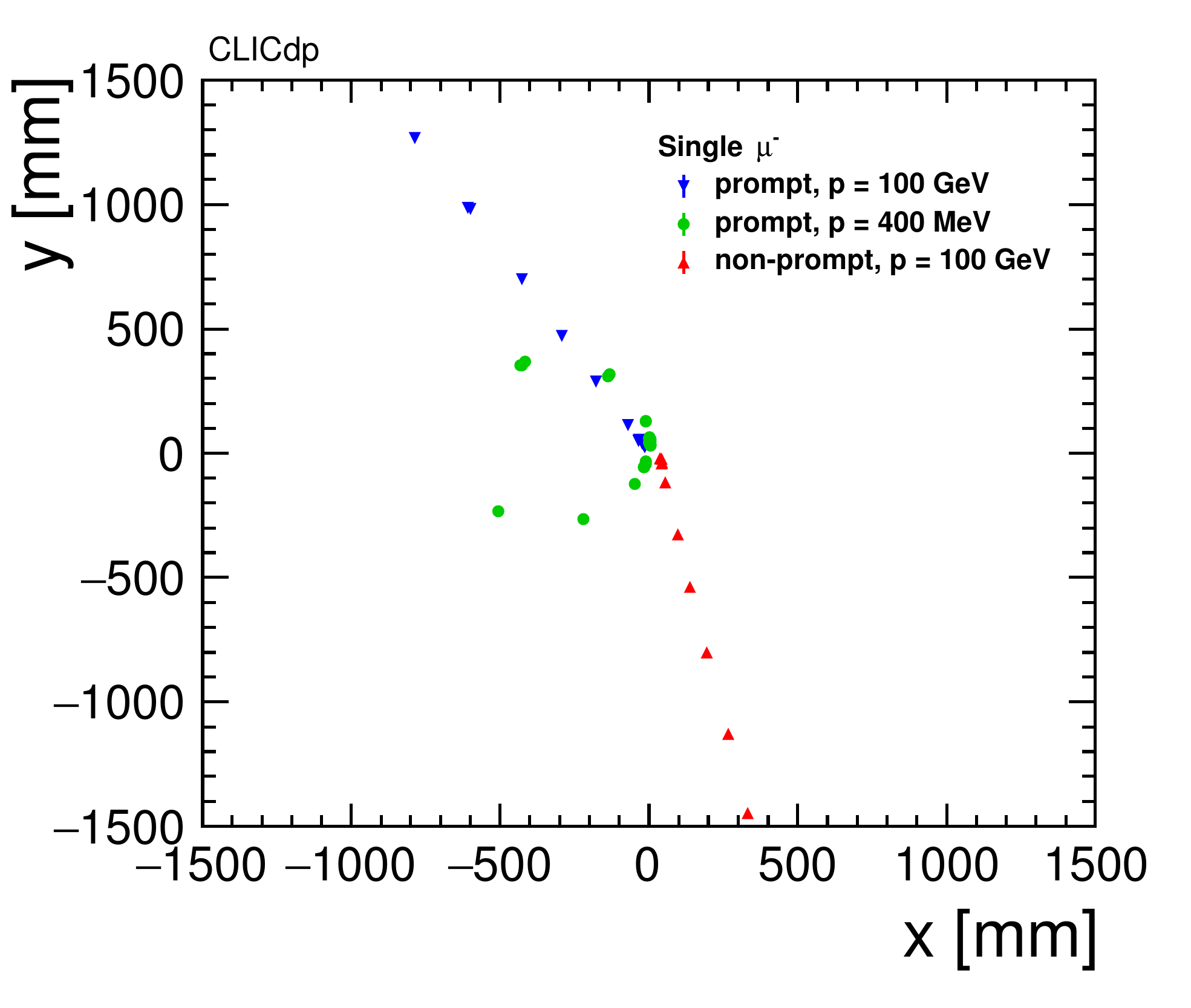}}
  \subfloat[]{\includegraphics[width=0.48\linewidth]{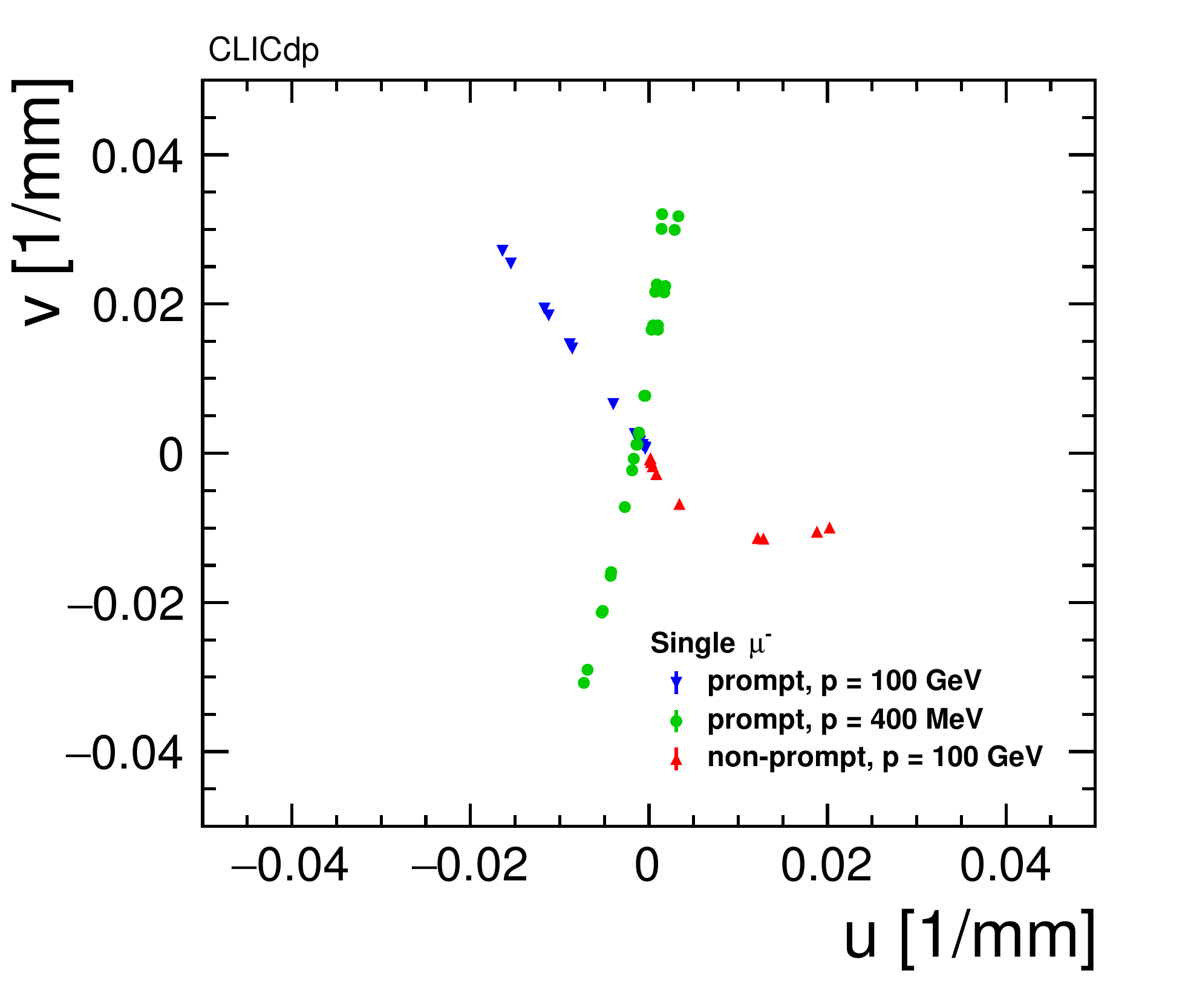}}
  \caption{Hits produced by three muon tracks: prompt (blue), prompt with low momentum (green) and non-prompt (red).
  Tracks are shown in the $(x,y)$ global coordinate system (a) and in the $(u,v)$ conformal coordinate system (b).}
\label{fig:conformal_mapping}
\end{figure}

\subsection{Track finding with the conformal tracking method}
\label{subsec:pattern_recognition}

In the conformal algorithm, point coordinates in global space $(x,y)$ are translated into the conformal space $(u,v)$.
The idea behind this coordinate transformation is that circles passing through the origin of a cartesian coordinate system $(x,y)$,
can be transformed into straight lines in a different coordinate system where the coordinates are defined as:
\begin{linenomath}
\begin{equation}
u = \frac{x}{x^{2}+y^{2}}, \qquad v = \frac{y}{x^{2}+y^{2}}.
\end{equation}
\end{linenomath}
This is demonstrated in~\cite{HANSROUL1988498} and is valid if the circle is passing through the origin.
In the conformal space $(u,v)$, the circle equation $(x-a)^{2} + (y-b)^{2} = r^{2}$ is translated into a straight line
\begin{linenomath}
\begin{equation}
v = -\frac{a}{b}u + \frac{1}{2b}.
\label{eq:straight_line}
\end{equation}
\end{linenomath}
Through the application of the conformal algorithm, 
curved tracks left by charged particles bent by a magnetic field can be reduced to a straight lines search.
The radial order of the hit positions is inverted in the conformal
space with respect to the global space: track
hits on the vertex detector layers correspond to larger $(u,v)$ coordinates than hits on the tracker layers.

In real measurements not all particle trajectories can be approximated by 
a straight line due to possible deviations caused by particles undergoing multiple scattering
or simply not originated at the origin of the $(x,y)$ plane, also known as \emph{displaced} or \emph{non-prompt} particles.
An example of each of these three cases is given in Figure~\ref{fig:conformal_mapping}
where hits belonging to three muon tracks are simulated using CLICdet.
To take these deviations systematically into account, pattern recognition in conformal space is performed via a cellular automaton (CA)~\cite{KISEL200685}.

In the CA, cells are defined as segments connecting two hits and the algorithm is based on the creation and extension of these cells.
The role of the CA in the conformal tracking is two-fold: firstly, it builds the so-called cellular track candidates 
and then, these cellular track candidates are extended to create full tracks and only those with the highest quality are selected.

\subsubsection{Building of cellular track candidates}
\label{subsubsec:build}

The hit collection given as input to the building step is referred to as the \emph{seeding collection}
and each hit in the collection is considered as a \emph{seed hit}, the starting point of a cellular track candidate.
From each seed hit, one or more \emph{seed cells} are created connecting the seed hit with its nearest neighbours.
The nearest neighbour is defined as a hit which is within a certain angular region and distance from the seed hit,
does not lie on the same detector layer, is located at a smaller conformal radius and is not already part of a track.
Each seed cell is then extended to \emph{virtual hits}, i.e. hits sitting on the prolongation in the radial direction 
of the cell, from which the search for nearest neighbours is repeated.
New cells are then created and attached to the seed cell if the neighbouring hits
do not lie on the same detector layer as the end point of the seed cell, have smaller conformal radius and had not been already included in a track.
The extension of the new cell is then repeated and more cells are created.
Each cell contains information about its start and end point as well as a \emph{weight} information. 
The weight of a cell indicates how many other cells are further connected to it. 
Each subsequent link increments the cell weight by one unit, such that the higher the weight, 
the higher the potential of the cell to make a track. 

Cellular track candidates are chains of cells, created by all cells compatible within a certain 
angle window starting from the highest weighted seed cell. A minimum number of hits for each candidate is required.
Among all the cellular track candidates only those with the best quality are kept.
This step is called \emph{clone treatement} and it assesses the candidate quality
using the length of the track and the two $\chi^{2}$ independently computed 
using a linear regression fit on the conformal space and on the $(s,z)$ plane, where $s$ is the coordinate along the helix arc segment.
Longer tracks are preferred, while the one with the best $\chi^{2}$ is chosen in case of equal length.
A further attempt to recover rejected cellular tracks is made by removing, one by one, each hit on the track, refitting and recomputing the normalised $\chi^{2}$. This allows one to keep good tracks that contain a spurious hit.
At the end of the process, all hits belonging to the created cellular tracks are marked as \emph{used}.

\subsubsection{Extension of cellular track candidates}
\label{subsubsec:extend}

The hit collection given as input of this CA step is used to extend the cellular track candidates created at the end of the building algorithm.
Cellular track candidates are extended in different ways according to their particle transverse momentum
estimated using the parameters extracted from the linear regression fit in the $(s,z)$ space.
All candidates with an estimated \pt above a threshold defined as external parameter are extended first, as they are easier to reconstruct.

In this case, the extension proceeds in a similar manner as described for the building step: 
end points of the previously formed cellular tracks are used as seed hits and a search for nearest neighbours in polar angle is performed.
The search is performed limiting the compatible hits in the successive detector layers.
New cellular tracks are created including every valid neighbour. Among those, 
the best track candidate is chosen based on the $\chi^{2}$ values from the linear regression fit in $(u,v)$ and $(s,z)$. 
If no hit is found on one layer, the cellular automaton proceeds on the subsequent layer.
The extension process is repeated until the last detector layer.

The extension algorithm then is performed for cellular track candidates whose estimated \pT is lower than the threshold.
To allow the reconstruction of very low-\pT tracks, such as particles looping in the tracker,
no specific nearest neighbour search is performed but
all unused hits which are not located on the other side of the detector in $z$ are considered for track extension.
In this part, the search for compatible hits is done including a quadratic term in the regression formula, 
in order to take into account deviations from a straight line.
At the end of the extension, all CA track candidates are sorted by \pT and the  hits included in the conformal tracks are marked as \emph{used}.

\subsubsection{Full conformal tracking chain}
\label{subsubsec:chain}

\begin{figure}[t]
  \centering
  \subfloat[]{\includegraphics[width=0.48\linewidth]{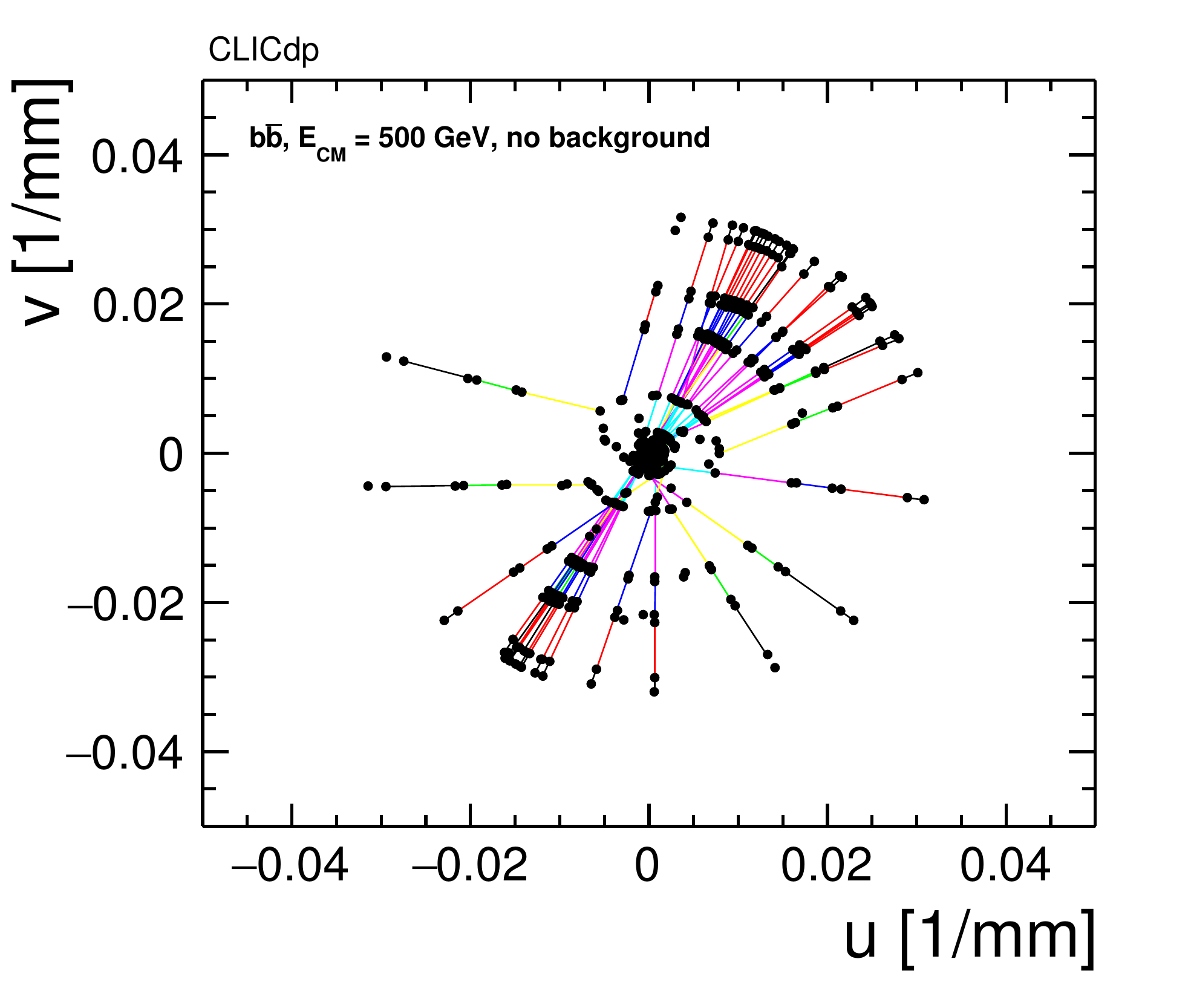}}
  \subfloat[]{\includegraphics[width=0.48\linewidth]{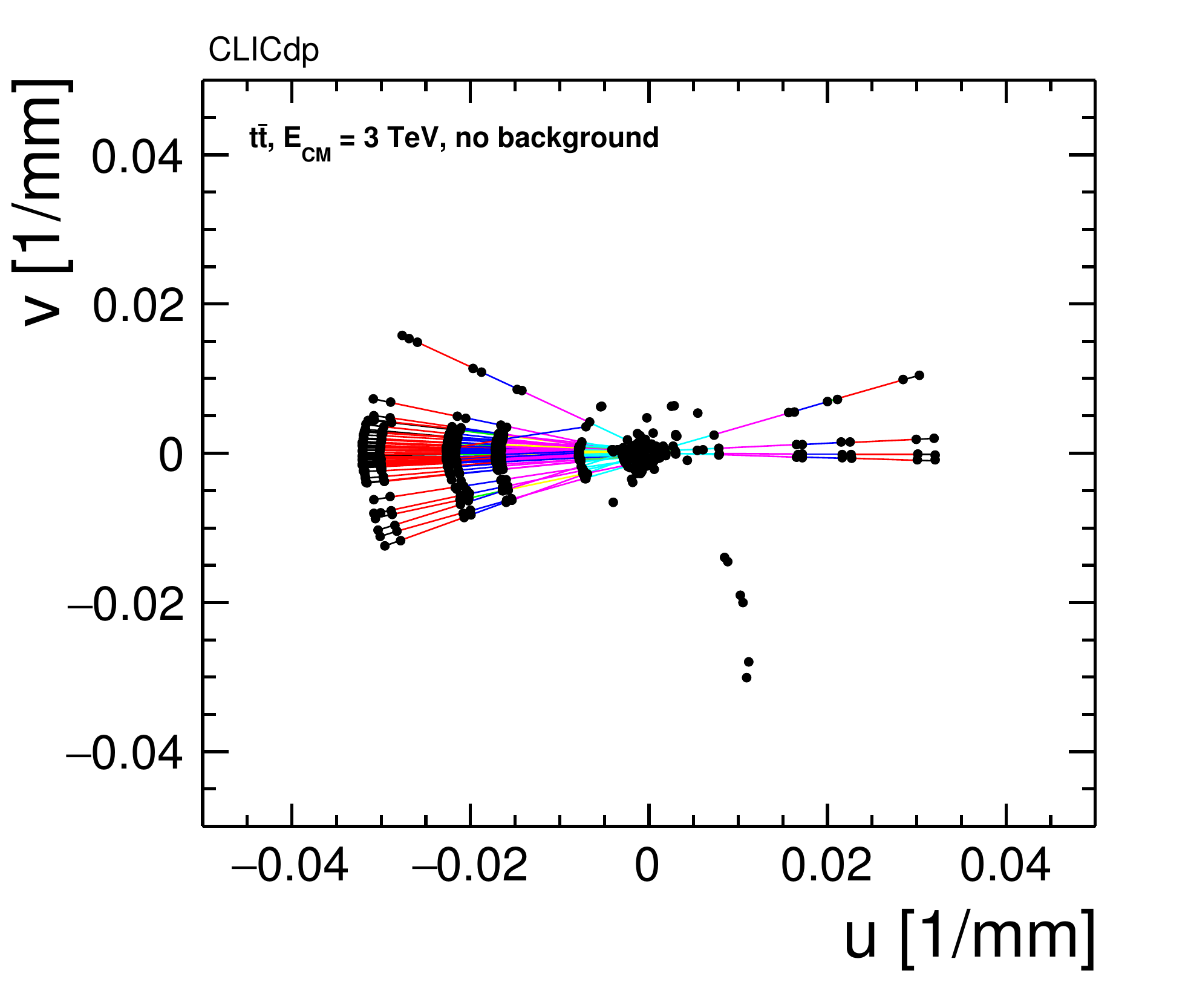}}
  \caption{Cellular track candidates reconstructed for a single \bb event at \SI{500}{\GeV} centre-of-mass energy (a) and
  for a single \ttbar event at \SI{3}{\TeV} (b).}
\label{fig:CAtracks}
\end{figure}

The two main CA algorithms are run on all reconstructed hits in several iterations with different sets of parameters.
The full chain used in the CLIC track reconstruction is shown in Table~\ref{tab:iterations}.
The first part aims to reconstruct the prompt tracks in the vertex detector: 
step 0 builds cellular track candidates using the hits in the vertex barrel as input, and
they are extended in step 1 using the hits in the vertex endcap.
The remaining hits in the combined vertex barrel and endcap detectors are then used 
to recover prompt tracks more difficult to find. 
This is done in two steps using tighter cuts first (step 2) and looser cuts afterwards (step 3).
All cellular track candidates reconstructed in steps 2 and 3 are then extended to the 
hits belonging to the tracker detectors in step 4.
The last part of the full chain focuses on the reconstruction of non-prompt particles (step 5)
using the whole tracking system. During this step the CA is run from the outermost layer of the tracker
back to the innermost layer of the vertex detector. 
Moreover, a quadratic term is included in the regression formula used to fit the cellular track candidates in conformal space.
%Given the overall looser requirements, step 5 is the most computationally demanding part of the whole pattern recognition chain. CPU time estimates per step are shown in \ref{subsec:cpu}.
%Two configurations are summarised in \ref{tab:iterations}, as even looser cuts can be used when reconstructing single isolated particles without any drawback in terms of computing time.% as it is the case for realistic topologies in high occupancy environments.
An example of all cellular track candidates reconstructed using the conformal tracking algorithm is given in
Figure~\ref{fig:CAtracks}.

\begin{table}
  \begin{center}
  \caption{Overview of the configuration for the different steps of the pattern recognition chain. The last column shows some of the parameters of relevance for the cellular automaton as used for CLICdet: the maximum angle between cells, the maximum cell length, the minimum number of hits on a track, the maximum $\chi^{2}$ for valid track candidates, and the \pT threshold used to discriminate between the two variations of the algorithm of track extension.}
  \label{tab:iterations}
  \begin{tabular}{cllccccc}
  \hline
  \hline
  \multirow{3}{*}{\textbf{Step}} & \multirow{3}{*}{\textbf{Algorithm}} & \multirow{3}{*}{\textbf{Hit collection}} & \multicolumn{5}{c}{\textbf{Parameters}}\\
  \cline{4-8}
  & & & max cell & max cell & min & $\chi^{2}$ & \pT threshold\\
  & & & angle [\SI{}{rad}] & length [\SI{}{\mm}$^{-1}$] & $N_{\textnormal{hits}}$  & cut  & [\SI{}{\GeV}]\\
  \hline
  0 & Building & Vertex Barrel & 0.005 & 0.02 & 4 & 100 & - \\ 
  1 & Extension & Vertex Endcap & 0.005 & 0.02 & 4 & 100 & 10 \\ 
  2 & Building & Vertex & 0.025 & 0.02 & 4 & 100 & - \\ 
  3 & Building & Vertex & 0.05 & 0.02 & 4 & 2000 & - \\ 
  4 & Extension & Tracker & 0.05 & 0.02 & 4 & 2000 & 1 \\ 
  5 & Building & Vertex \& Tracker & 0.05 & 0.015 & 5 & 1000 & - \\ 
  % 5 \footnotesize{(isolated)} & Building & Vertex \& Tracker & 0.1 & 0.015 & 5 & 1000 & - \\
  \hline
  \hline
  \end{tabular}
  \end{center}
\end{table}

\subsection{Track fitting and selection}
\label{subsec:track_fitting}

The track fit in CLICdet consists of a Kalman filter and smoother in the global $(x,y)$ coordinate space~\cite{Fruehwirth:KF}.
The fit is initialised with the parameters obtained by a simple helix fit to three hits of the track -- typically first, middle, and last hit.
The Kalman filter and smoother is then run as implemented in the KalLib package described in~\cite{ILCsoft:KF}.
If the fit fails, the Kalman filter and smoothing procedure is tried again backwards.

In the track selection, some cuts are applied to the track candidates to filter 
out the tracks that contain many spurious hits or share a number of hits with other tracks. 
In order to do that, the clone treatment described in Section~\ref{subsubsec:build} 
is performed on the fitted track collection. 
The $\chi^{2}$ used to determine the highest quality track is the one calculated 
at the end of the Kalman filter and smoother procedure. 
Moreover, only tracks with a number of hits larger or equal to three is kept
to maximise the track efficiency.

\section{Performance}
\label{sec:performance}

In this section, the performance of the track reconstruction for the CLICdet
experiment is investigated for simulated isolated muons and realistic topologies in a high occupancy environment.
% Finally, \ref{subsec:cpu} provides an analysis of the CPU time required for the different steps of the tracking algorithm.

The tracking efficiency is defined as the ratio of \emph{pure} tracks to the 
number of reconstructable particles, i.e. simulated particles which are stable, and have $\pT > \SI{100}{\MeV}$, $|\cos(\theta)| < 0.99$, 
and at least 4 hits on different layers.
Pure tracks are defined as reconstructed tracks which have at least 75\% of their hits associated to
a single reconstructable particle.
The efficiency is presented as a function of various parameters of the simulated particles.

The fake rate is defined as the ratio of \emph{fake} tracks to the total number of reconstructed tracks.
Fake tracks are defined as reconstructed tracks which have less then 75\% of their hits associated to one reconstructable particle.
The fake rate is presented as a function of various parameters of the reconstructed track.

The resolution for a specific track parameter is estimated from the track parameter residuals, 
defined as the difference between the reconstructed-track parameter and the simulated-particle parameter.
The distribution of residuals is fitted with a Gaussian function, whose standard deviation defines the resolution.

\subsection{Results for isolated muons}
\label{sec:results_muons}

\begin{figure}[tb]
  \centering
  \subfloat[]{\includegraphics[width=0.48\linewidth]{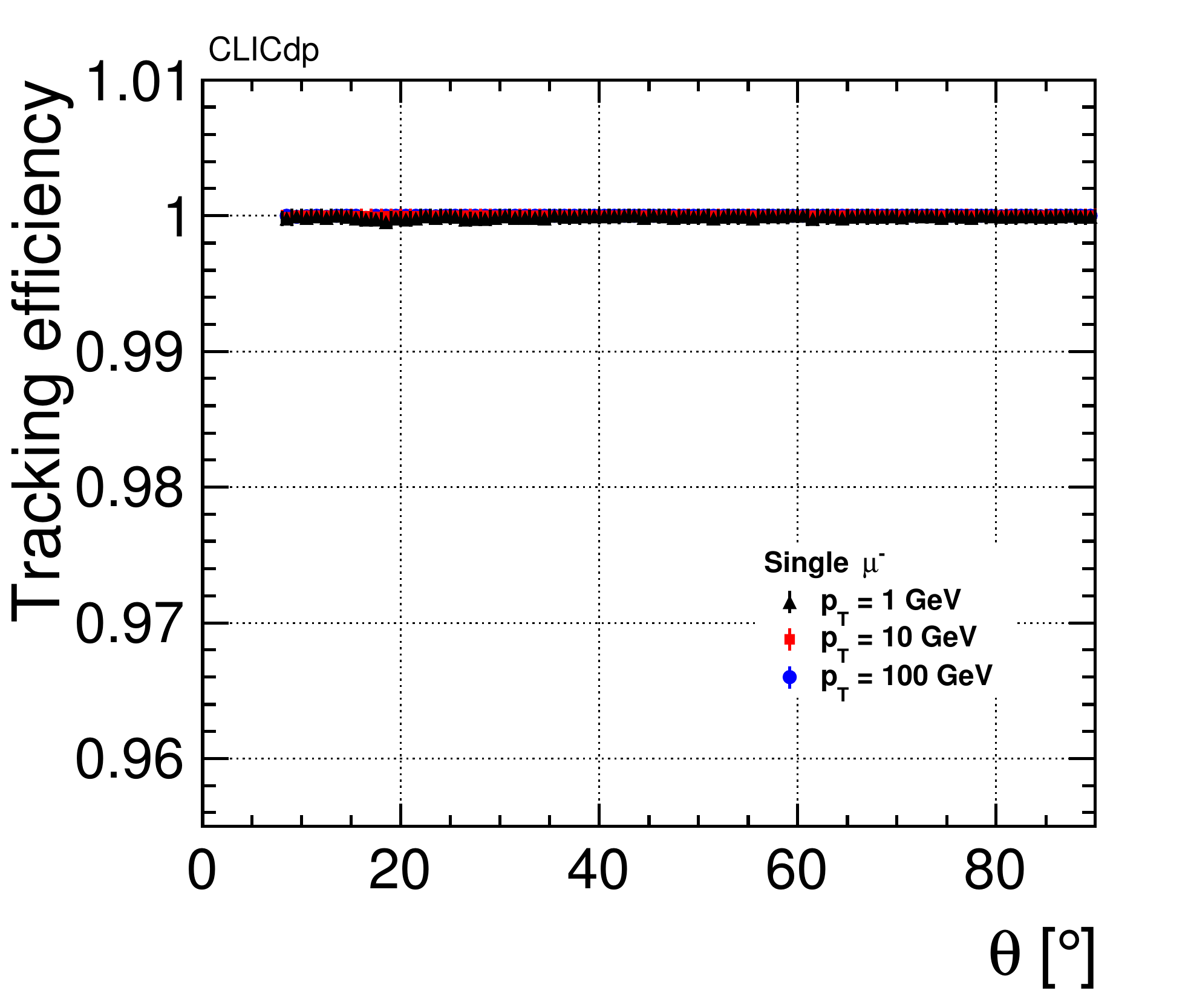}}
  \subfloat[]{\includegraphics[width=0.48\linewidth]{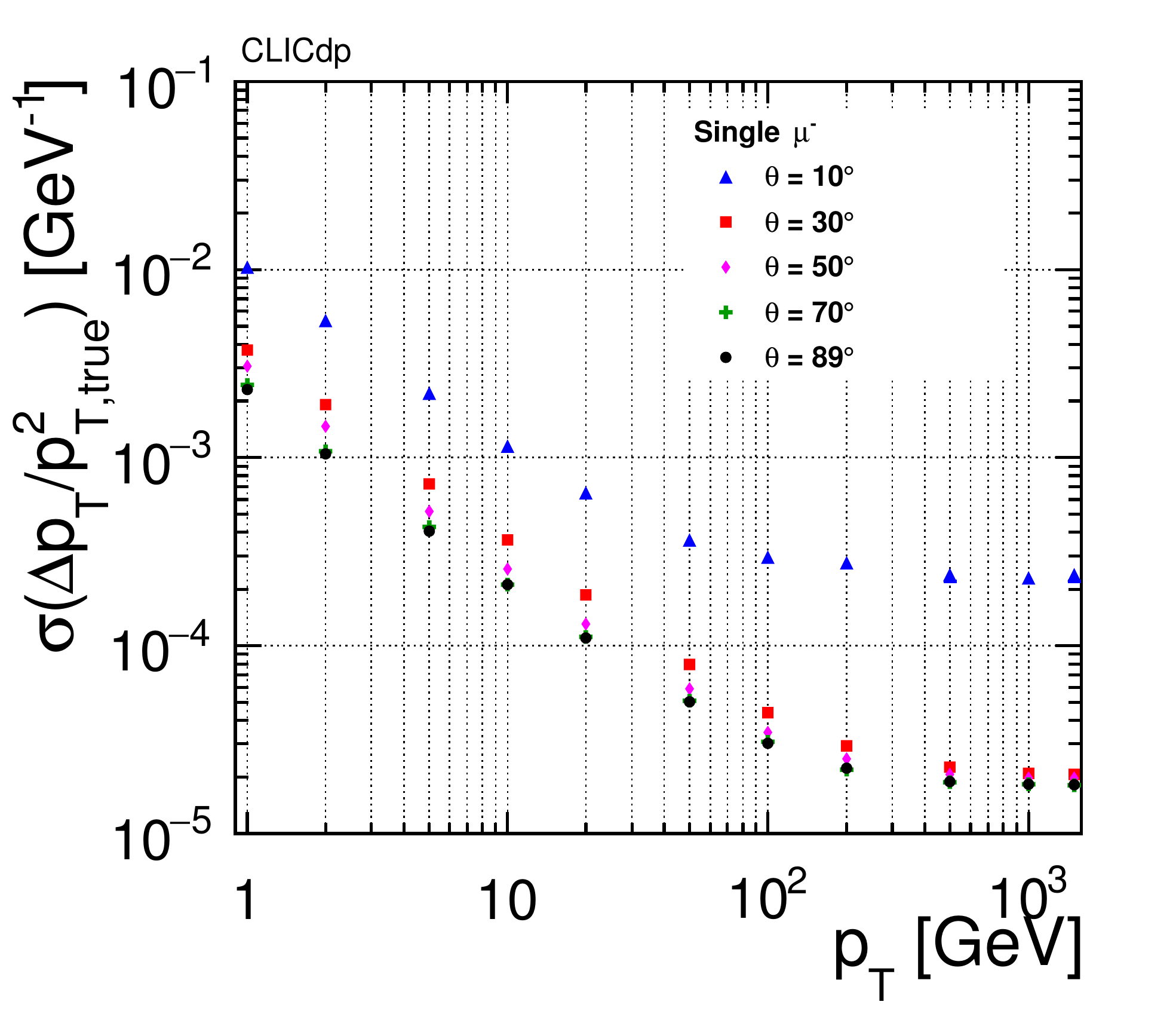}}
  \caption{(a) Track reconstruction efficiency  as a function of polar angle $\theta$ for \pT = \SI{1}{\GeV}, \SI{10}{\GeV}, and \SI{100}{\GeV}
  for isolated muons reconstructed with the CLICdet tracker.
  (b) Transverse momentum resolution for isolated muons with polar angle $\theta$ = \ang{10}, \ang{30}, \ang{50}, \ang{70} and \ang{89} 
  as a function of \pT.}
  \label{fig:muons_perf}
\end{figure}

The tracking efficiency for single isolated muons is shown in Figure~\ref{fig:muons_perf}(a)
as a function of the polar angle of the simulated particle.
The tracking is fully efficient in the entire tracker acceptance and for any transverse momentum~\cite{CLICperf}, while 
the fake rate is negligible.

%To defer the efficiency for displaced particles, a dedicated study was performed 
%with isolated muons produced with a displacement up to \SI{60}{\cm} in the $y$ coordinate and with a polar angle uniformly distributed in a \ang{10} cone around the $y$ axis.
%The result is shown in Figure~\ref{fig:muons_vtx_reso}(a) for muons with momenta of \SI{1}{\GeV}, \SI{10}{\GeV}, and \SI{100}{\GeV} as a function of the particle production vertex radius. 
%The efficiency is close to 100\% for the higher-energy particles sample, up to a production radius of \SI{35}{\cm}
%as allowed by the minimum number of layers required in the conformal tracking.
%Muons with momenta of \SI{1}{\GeV} are efficiently reconstructed within the vertex detector region, i.e. with radius smaller than \SI{6}{\cm}. 

The resolutions of the following track parameters were studied using simulated isolated muons~\cite{CLICperf}: 
transverse and longitudinal impact parameter, azimuthal and polar angle, and transverse momentum.
High-momentum tracks fulfil the requirements on the impact parameter and transverse momentum
set by the physics programme at CLIC as described in Section~\ref{sec:clic}.
As expected, all track parameter resolutions deteriorate in the forward region
and for low-momentum tracks due to multiple scattering.
Moreover, for high-momentum tracks resolutions of the impact parameters achieve the limit foreseen by the
assumption on the single point resolution in the vertex detector.
As an example, Figure~\ref{fig:muons_perf}(b) shows the resolution of the transverse momentum parameter
as a function of the \pT of the simulated muon produced at $\theta$ = \ang{10}, \ang{30}, \ang{50}, \ang{70} and \ang{89}. 

%\begin{figure}[!htb]
%  \centering
%  \subfloat[]{\includegraphics[width=0.48\linewidth]{figures/muons_eff_vs_theta_minNhits3}}
%  \subfloat[]{\includegraphics[width=0.48\linewidth]{figures/muons_eff_vs_pT_minNhits3}}
%  \caption{Track reconstruction efficiency  as a function of polar angle $\theta$ for \pT = \SI{1}{\GeV}, \SI{10}{\GeV}, and \SI{100}{\GeV} (a) 
%  and as a function of \pT for $\theta$ = \ang{10}, \ang{30}, and \ang{89} (b) for isolated muons reconstructed with the CLICdet tracker.}
%  \label{fig:muons_eff}
%\end{figure}
%
%\begin{figure}[!htb]
%  \centering
%  \subfloat[]{\includegraphics[width=0.48\linewidth]{figures/muons_eff_vs_vtx}}
%  \subfloat[]{\includegraphics[width=0.48\linewidth]{figures/muons_pt_res_pt.eps}}
%  \caption{(a) Track reconstruction efficiency as a function of the production vertex radius for \pT = \SI{1}{\GeV}, \SI{10}{\GeV}, and \SI{100}{\GeV}
%  for isolated muons reconstructed with the CLICdet tracker.\\
%  (b)Transverse momentum resolution for isolated muons with polar angle $\theta$ = \ang{10}, \ang{30}, \ang{50}, \ang{70} and \ang{89} as a function of \pT.}
%  \label{fig:muons_vtx_reso}
%\end{figure}

\subsection{Results for complex events}
\label{sec:results_ttbar}

The robustness of the track reconstruction at CLIC is estimated using more complex events
in a realistic CLIC environment. Simulated \epem$\rightarrow$ \ttbar events at \SI{3}{\TeV} centre-of-mass energy
with and without the \gghadron background expected at the \SI{3}{\TeV} energy stage of CLIC are used for this study.

\begin{figure}[bt]
  \centering
  \subfloat[]{\includegraphics[width=0.48\linewidth]{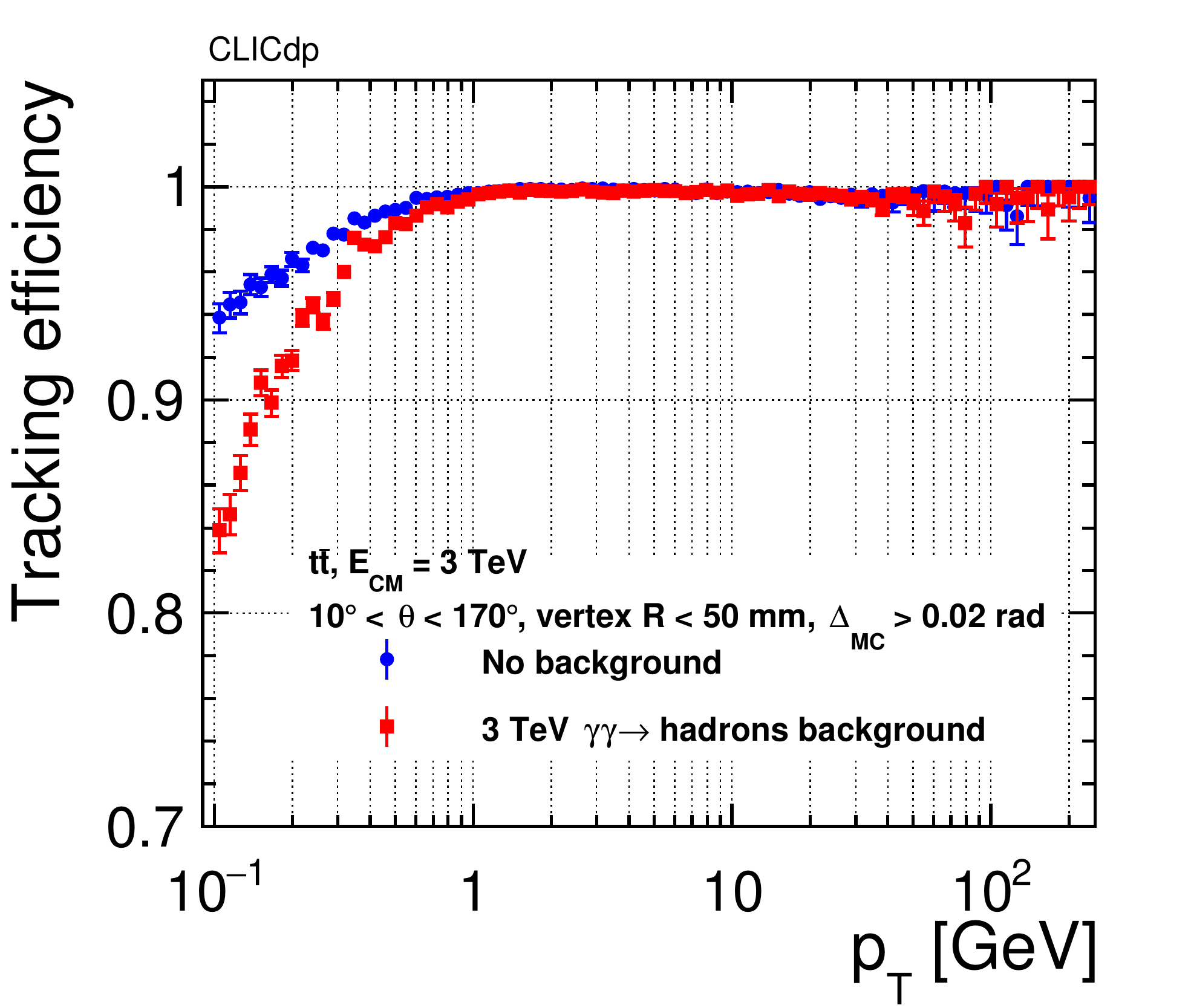}}
  \subfloat[]{\includegraphics[width=0.48\linewidth]{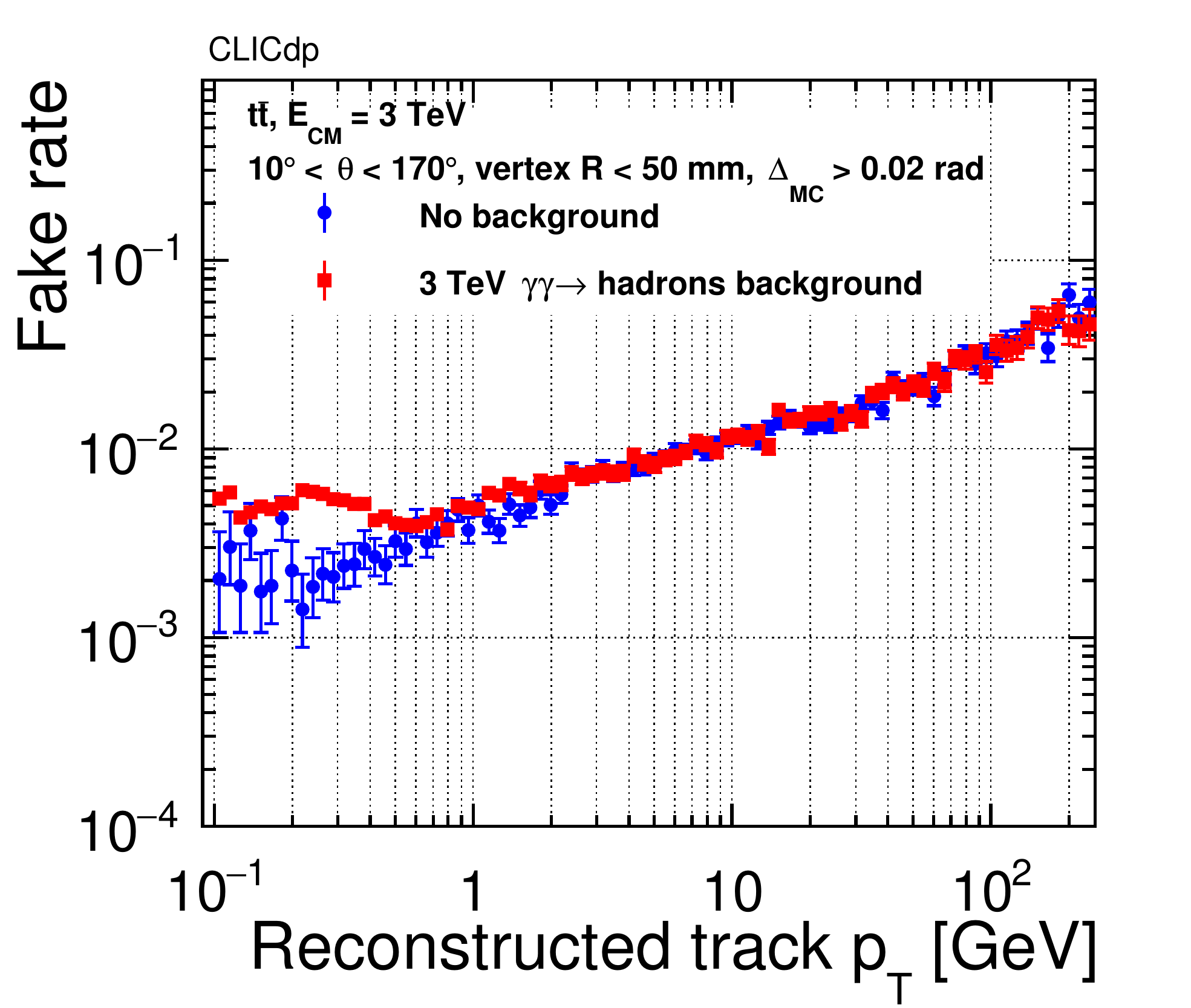}}
  \caption{Tracking efficiency (a) and fake rate (b) as a function of the \pT of the simulated particle and of the reconstructed track, respectively.
  \epem$\rightarrow$ \ttbar events simulated at \SI{3}{\TeV} centre-of-mass energy are used with and without the corresponding overlay of 30 BX of 
  \gghad background expected at the \SI{3}{\TeV} energy stage of CLIC.}
  \label{fig:ttbar}
\end{figure}

In Figure~\ref{fig:ttbar} the efficiency and fake rate are shown as a function of the \pT of the simulated particle
and of the reconstructed track, respectively. 
The tracking is fully efficient for simulated particles with $\pT > \SI{1}{\GeV}$ and the efficiency is
still well above 90\% down to a \pT approximately of \SI{200}{\MeV}. 
%The tracking is also fully efficient for
%particles produced in the entire tracking volume ($\theta > \ang{20}$).
Both efficiency and fake rate are almost unaffected by the \gghad background expected at the \SI{3}{\TeV} energy stage of CLIC.

\begin{figure}[!htb]
  \centering
  \subfloat[]{\includegraphics[width=0.48\linewidth]{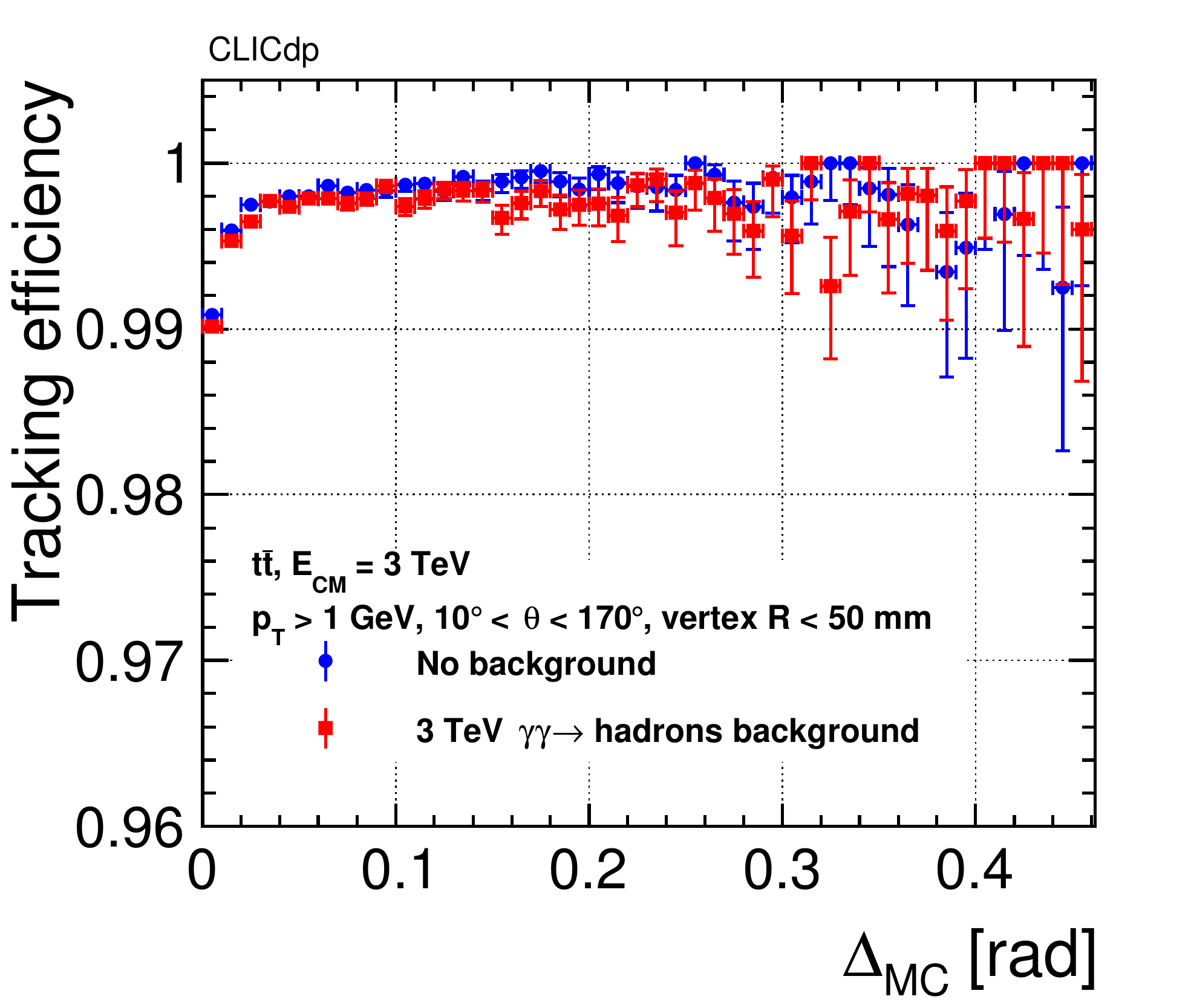}}
  \subfloat[]{\includegraphics[width=0.48\linewidth]{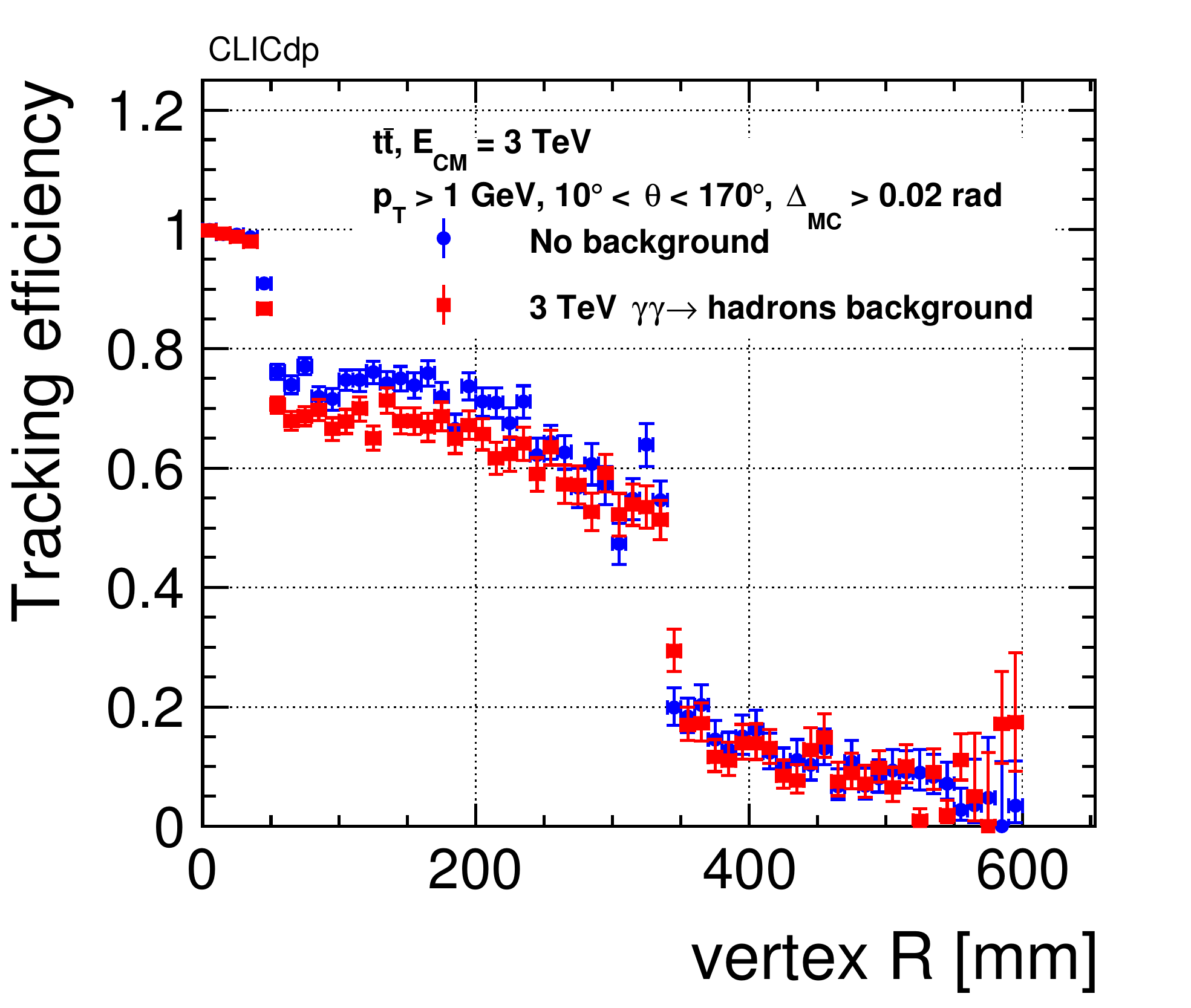}}
  \caption{Tracking efficiency as a function of particle proximity (a) and of the particle production vertex radius (b).
  \epem$\rightarrow$ \ttbar events simulated at \SI{3}{\TeV} centre-of-mass energy are used with and without the corresponding overlay of 30 BX of
  \gghad background expected at the \SI{3}{\TeV} energy stage of CLIC.}
  \label{fig:ttbar_2}
\end{figure}

Results were also studied as a function of the smallest distance between the particle associated to the track and any other particle
and as a function of particle production vertex radius. The efficiency plots are shown in Figure~\ref{fig:ttbar_2}.
The maximum efficiency loss for tracks produced in the vertex detector amounts to only 1\%
for $\Delta_{\textnormal{MC}} < \SI{0.02}{\rad}$.
The tracking is fully efficient provided that the particles are produced within the vertex detector.
%while the edge of the vertex detector was found to be a particularly difficult region to reconstruct tracks --
%the efficiency drops and the fake rate presents a peak of around 10\%. 

Thanks to the conformal tracking modularity,
the track reconstruction in CLIC was successfully adapted also to the CLD detector at FCC-ee
which not only has a different sub-detector design
and magnetic field magnitude but also includes different background conditions~\cite{Leogrande:2630512,Benedikt:2651299}.

\subsection{CPU execution time}
\label{subsec:cpu}

Like in many other particle physics experiments, track reconstruction is one of the most time consuming parts of the event reconstruction.
A first assessment was performed using the mean CPU time of ten \ttbar events at \SI{3}{\TeV} centre-of-mass energy
without and with the overlay of \gghad background expected at the \SI{3}{\TeV} energy stage of CLIC.
If the background is not included, most of the CPU time is spent in the fitting procedure, otherwise
the last CA building step (step~5) is the most time consuming part.

\section{Conclusions}
\label{sec:summary}

The conformal tracking presented in this paper is a new pattern recognition technique 
for track finding in the context of future electron--positron colliders.
It uses a cellular automaton algorithm to find efficiently prompt and non-prompt tracks in conformal space
and it has been demonstrated that it reconstructs tracks successfully also in the most challenging low-\pt regime.
Its robustness against the beam-induced backgrounds was proven
using more complex events for the highest energy stage of CLIC.
Moreover, some preliminary studies in terms of CPU computing time were also presented.
Its use in the Particle Flow Analysis foreseen for particle reconstruction 
at future electron--positron colliders is essential~\cite{CLICperf} and will trigger further improvements in the near future.

%%%%%%%%%%%%%%%%%%%%%%%%%%%%%%%%%%%%%%%%%%%%%%%%%%%%%%%%%%%%%%%%%%%%%%%%%%%

%%%%%%%%%%%%%%%%%%%%%%%%%%%%%%%%%%%%%%%%%%%%%%%%%%%%%%%%%%%%%%%%%%%%%%%%%%%

\end{document}